\documentclass[useAMS,usenatbib]{mn2e}

\newcommand{\be}{\begin{equation}}
\newcommand{\ee}{\end{equation}}
\newcommand{\ba}{\begin{eqnarray}}
\newcommand{\ea}{\end{eqnarray}}

\def\simless{\mathbin{\lower 2.5pt\hbox
   {$\rlap{\raise 4.5pt\hbox{$\char'074$}}\mathchar"7218$}}}
\def\simgrt{\mathbin{\lower 2.5pt\hbox
   {$\rlap{\raise 4.5pt\hbox{$\char'076$}}\mathchar"7218$}}}
\include{epsf}

\title[H-ATLAS: Modelling the first lenses]
{{\it Herschel}\thanks{{\it Herschel} is an ESA space observatory
      with science instruments provided by European-led Principal
      Investigator consortia and with important participation from
      NASA.}-ATLAS: Modelling the first strong gravitational lenses}
\author[S. Dye et al.]
{\parbox{\textwidth}{S. Dye,$^{1}$\thanks{E-mail:simon.dye@nottingham.ac.uk}
M. Negrello$^{2}$,
R. Hopwood$^{3}$,
J. W. Nightingale$^{1}$, 
R. S. Bussmann$^{4,12}$,
S. Amber$^{5}$,
N. Bourne$^{1,6}$,
A. Cooray$^{7}$, 
A. Dariush$^{3}$, 
L. Dunne$^{8}$, 
S. A. Eales$^{9}$,
J. Gonzalez-Nuevo$^{10}$,
E. Ibar$^{11}$, 
R. J. Ivison$^{6}$,
S. Maddox$^{8}$, 
E. Valiante$^{9}$, 
M. Smith$^{9}$}
\vspace{4mm}\\
$^{1}$School of Physics and Astronomy, Nottingham University,
University Park, Nottingham, NG7 2RD, UK\\
$^{2}$INAF, Osservatorio Astronomico di Padova, Vicolo Osservatorio 5,
I-35122 Padova, Italy\\
$^{3}$Astrophysics Group, Imperial College London, Blackett Laboratory,
Prince Consort Road, London, SW7 2AZ, UK\\
$^{4}$Harvard Smithsonian Center for Astrophysics, 60 Garden St.,
Cambridge, MA 02138, USA\\
$^{5}$Department of Physical Sciences, The Open University, Walton Hall,
Milton Keynes, MK7 6AA, U.K.\\
$^{6}$Institute for Astronomy, Royal Observatory Edinburgh, 
Blackford Hill, Edinburgh, EH9 3HJ, UK.\\
$^{7}$Astronomy Department, California Institute of Technology,
MC 249-17, 1200 East California Boulevard, Pasadena, CA
91125, USA. \\
$^{8}$Department of Physics and Astronomy, University of Canterbury,
Private Bag 4800, Christchurch, 8140, New Zealand\\
$^{9}$School of Physics and Astronomy, Cardiff University, The Parade, 
Cardiff, CF24 3AA, UK.\\
$^{10}$Instituto de F\'{i} sica de Cantabria (CSIC-UC), Av. los Castros
s/n, 39005 Santander, Spain.\\
$^{11}$Instituto de Astrof\'{i}sica, Facultad de F\'{i}sica,
Pontificia Universidad Cat\'{o}lica de Chile, Casilla 306,
Santiago 22, Chile.\\
$^{12}$Department of Astronomy, Space Science Building, Cornell University,
Ithaca, NY, 14853-6801.\\
}

\begin{document}

\date{}

\pagerange{\pageref{firstpage}--\pageref{lastpage}} 
\pubyear{2014}

\maketitle

\label{firstpage}

\begin{abstract}
  We have determined the mass-density radial profiles of the first
  five strong gravitational lens systems discovered by the Herschel
  Astrophysical Terahertz Large Area Survey (H-ATLAS). We present an
  enhancement of the semi-linear lens inversion method of Warren \&
  Dye which allows simultaneous reconstruction of several different
  wavebands and apply this to dual-band imaging of the lenses acquired
  with the Hubble Space Telescope. The five systems analysed here have
  lens redshifts which span a range, $0.22 \leq z \leq 0.94$. Our
  findings are consistent with other studies by concluding that: 1)
  the logarithmic slope of the total mass density profile steepens
  with decreasing redshift; 2) the slope is positively correlated with
  the average total projected mass density of the lens contained
  within half the effective radius and negatively correlated with the
  effective radius; 3) the fraction of dark matter contained within
  half the effective radius increases with increasing effective
  radius and increases with redshift.
\end{abstract}


\section{Introduction}

Early type galaxies, despite being relatively well studied, continue
to challenge our complete understanding of their formation and
evolution.  Current unanswered questions include quantifying the role
of mergers in their evolution
\citep[e.g.,][]{dokkum99,khochfar03,bell06,hilz13}, reliably
determining their stellar build up and reconciling this with
downsizing \citep[e.g.,][]{thomas05,maraston09,tojeiro12},
understanding the evolution of the upper end of the mass function
\citep[e.g., Bundy et al. 2005, 2007;][]{derwel09,hopkins10} and
identifying the process(es) by which the high redshift population
becomes dramatically less compact at low redshifts
\citep[e.g.,][]{daddi05,trujillo06,dokkum08,lani13}.  Regarding this
last point, mergers have been suggested as the cause of the effect,
but there is much disagreement \citep[e.g.,][]{hopkins10,oser12}.
Notwithstanding these unknowns, it seems likely that common formation
and evolution mechanisms are at play, given the tightness of observed
relationships such as the fundamental plane, the correlation of black
hole mass with central velocity dispersion and the near-isothermality
of total mass density profiles.

Regarding the measurement of density profiles, this has recently
become a very active pursuit within the field, motivated by the many
scientific applications made possible. These applications include the
provision of an observational benchmark for simulations of large scale
structure formation, constraining the initial mass function (IMF) by
comparing with stellar synthesis masses 
\citep[e.g.,][]{barnabe13,treu10,auger10}, determining the Hubble
constant from gravitational lens time delays \citep[e.g.][see also the
review by Jackson 2007]{tewes13} and cosmography \citep[see][and
  references therein]{treu10b}. Other novel applications include using
density profiles of strongly lensed systems embedded within a cluster
or group environment as a direct probe of the larger scale
gravitational potential \citep[e.g.,][]{dye07b,limousin10} and making
predictions of the self-annihilation signal of dark matter to guide
annihilation detection experiments \citep[e.g.,][]{walker11}.

A debate that continues to be re-kindled is the issue of whether
density profiles are cored, whereby the density tends to a constant
value towards small radii or whether they are cuspy, whereby the
density continues to increase as a radial power-law. 
To the advocates of cuspy profiles, the debate is over
the extent to which they are cuspy, i.e. the exponent of
the radial power-law. The motivation that drives these studies
originates from comparing observed density profiles with those
predicted by N-body simulations of large scale structure formation.

Early simulations favoured cuspy profiles that are typically steeper
than those inferred from observations \citep[see, for example,][and
  citations to this work]{blok02}.  However, these early studies were
largely based on pure dark matter simulations which ignored the
effects of baryons. Accordingly, the observational data concentrated
on dwarf galaxies where baryons behave more like test particles in a
dominating dark matter potential.  Recent work by \citet{cole12} shows
that cuspy dark matter profiles in simulated dwarf spheroidals results
in stronger dynamical friction causing globular clusters to fall into
the centres on a dynamical time scale, in contrast to what is observed
in these systems.

Simulations of large scale structure are now beginning to incorporate
baryons, but this is a highly complex task, fraught with many
complicating factors such as black hole accretion and their subsequent
feedback \citep[e.g.,][]{croton06,ciotti09,bryan13}, feedback from
supernovae and cooling \citep[see, for example,][]{duffy10,newton13}. 
Understanding the interplay of baryons and dark matter is not only
essential to a full comprehension of the formation and evolution of
galaxies, but can also shed light on the properties of the dark matter
itself, such as constraining the self-interaction cross-section of
dark matter \citep{spergel00,loeb11}. In this regard, \citet{lovell12}
find that the velocity profiles of satellite galaxies around the Milky
Way are considerably better matched by warm dark matter density
profiles than cold dark matter density profiles.

Improving the quality of observations of density profiles of galaxies,
particularly the more poorly understood early types, therefore
provides a much needed benchmark to assist in discrimination of the
many different scenarios describing their history. Gravitational
lensing offers a very powerful and yet conceptually simple approach to
achieving this, independent of assumptions about the kinematical state
of the deflecting mass. Whilst weak galaxy-galaxy lensing can be used
to constrain density profiles, this, by necessity, must be conducted
in a statistical sense \citep[e.g.,][]{velander13} and provides
measurements of the density profile on larger scales where the dark
matter dominates. Conversely, strong galaxy-galaxy lensing is
applicable on a per-galaxy basis and is sensitive to density profiles
on small scales where the poorly constrained baryon physics is more
dominant.

Early type galaxies have a higher average lensing cross-section than
disk galaxies \citep[e.g.,][]{maoz03} and hence strong galaxy-galaxy
lens samples tend to harbour significantly more early than late type
lenses. In this way, such samples provide a perfect opportunity to
gain unique insights into the formation of early types, a fact that
has inspired the culmination of several different lens samples to
date. Recent strong galaxy-galaxy lens samples include the Sloan
Lens ACS (SLACS) survey \citep{bolton06} with 85 lenses out to a
redshift of $z\simeq 0.4$ \citep[median redshift $\simeq$
0.2;][]{auger09}, the Strong Lensing Legacy Survey
\citep[SL2S;][]{cabanac07} with a final total of 36 lenses in the
range $0.2 \leq z \leq 0.8$ \citep[median redshift $\simeq
0.5$;][]{sonnenfeld13}, the Sloan WFC Edge-on Late-type Lens Survey
\citep[SWELLS;][]{treu11} with 20 disk galaxy lenses at $z \simless
0.2$ \citep{dutton13} and 20 lenses at $0.4 \leq z \leq 0.7$
identified in the Baryon Oscillation Spectroscopic Survey
\cite[BOSS;][] {eisenstein11} which constitute the BOSS
Emission-Line Lens Survey \citep[BELLS;][]{brownstein12}.
Regardless of their selection as lensing galaxies, both the SWELLS
and the SLACS samples are found to be statistically consistent
with being drawn at random from their parent un-lensed samples
with the same mass and redshift distributions
\citep{bolton08,treu11}.  This fortifies the role of lens samples
in their aforementioned applications.

Analysis of these existing lens surveys has already enabled some
interesting insights into galaxy evolution. For example, 
the fraction of dark matter within half the
effective radius of early types increases with galaxy size and mass
\citep{auger10b,ruff11}.  Another intriguing result which has direct
consequences for simulations of large scale structure is that the
total density profile of early type galaxies appears to steepen with
decreasing redshift \citep{ruff11,bolton12,sonnenfeld13} although the
degree to which this occurs is currently in disagreement.

In this paper, we present modelling of the first five strong
galaxy-galaxy lens systems identified in the Herschel Astrophysical
Terahertz Large Area Survey \citep[H-ATLAS;][]{eales10}.  The H-ATLAS
is a large area survey ($\sim 550$\, deg$^2$) conducted in five
passbands in the sub-millimetre (submm) wavelength range $100\,\mu$m
$\simless \lambda \simless 500\, \mu$m using the Herschel Space
Observatory \cite{pilbratt10}. Being a submm survey, the negative
K-correction afforded by submm galaxies means that lensed sources are
much more readily detected out to significantly greater redshifts than
surveys conducted at optical wavelengths. Increasing the distance of a
source increases the probability of it being lensed by intervening
matter. Combining this fact with the large areal coverage of H-ATLAS
results in an anticipated sample of hundreds of strong galaxy-galaxy
lenses \citep{negrello10,gonzalez12}. 

Such lenses also have the advantage that their submm emission is
unaffected by any dust in the lens which means that a clean view of
the source is obtained. This proves particularly important when
reconstructing high resolution surface brightness maps of the high
redshift lensed source for morphological studies.  Furthermore, submm
galaxy number counts are steep so that strongly lensed galaxies can be
straightforwardly identified with simple flux selection criteria. This
simple technique has also been applied by the HERschel Multi-tiered
Extragalactic Survey \cite[HerMES; Oliver et al. 2012. See] [for an
  account of the lensing aspects of this survey]{wardlow13} and the
survey carried out at mm wavelengths by the South Pole Telescope
\citep{carlstrom11,vieira13}.

The ultimate size of the H-ATLAS sample is an obvious advantage in
terms of improving statistical uncertainties. However, another far
more compelling benefit that arises from the negative K-correction in
the submm is that higher redshift lensed sources are more likely to be
lensed by higher redshift lenses. From an evolutionary point of view,
this brings about an increase in the period over which transformations
in density profiles can be determined, back to earlier times in the
Universe's history when the rate of galaxy evolution was stronger
\citep[for a theoretical perspective, see, for e.g.,][]{schaye10}.  In
addition to this, almost all submm galaxies have extended structure on
the scales of typical galaxy lens caustics so that their lensed images
comprise extended arcs and ring-like structures. As demonstrated by
\citet{dye05} in application of the semi-linear inversion (SLI)
algorithm \citep{warren03}, extended structure in lensed images allows
stronger constraints to be placed on the density profile of the
lensing galaxy.

In this paper, we apply an enhanced version of the SLI method which
allows multiple datasets observed at different wavelengths to be
simultaneously reconstructed with the same lens model. We apply this
to Hubble Space Telescope (HST) images acquired in both the F110W and
F160W filters of each of the five lenses identified in the
14.4\,deg$^2$ data released by the H-ATLAS consortium as science
demonstration phase (SDP) data.

The layout of this paper is as follows: Section \ref{sec_data}
outlines the data. In Section \ref{sec_method} we describe the
methodology of the lens modelling, including a description of the
enhanced SLI method.  Section \ref{sec_results} presents the results
and we summarise the findings of this work in Section
\ref{sec_summary}.  Throughout this paper, we assume the following
cosmological parameters; ${\rm H}_0=67\,{\rm km\,s}^{-1}\,{\rm
  Mpc}^{-1}$, $\Omega_m=0.32$, $\Omega_{\Lambda}=0.68$ \citep{planck13}.

\section{Data}
\label{sec_data}

The data analysed in this work are more thoroughly described in a
companion paper (Negrello et al., 2013, hereafter denoted N13) but we
include the pertinent details here for completeness.

The HST observations were carried out in April 2011 in Cycle 18 under
proposal 12194 (PI Negrello) using the Wide Field Camera 3 (WFC3). Two
orbits were allocated per target with at least three quarters of the
total exposure time per target of 5130s acquired in the F160W filter
and the remainder in the F110W filter.  Images were reduced using the
IRAF {\tt MultiDrizzle} package and resampled to a pixel scale of
0.064'', half the intrinsic pixel scale of the WFC3.

The lens galaxy flux and lensed background source image in each system
were then simultaneously fitted with smooth profiles using the {\tt
  GALFIT} software \citep{peng02} and the lens profiles subtracted to
leave the lensed image. These lens-subtracted images, along with their
corresponding noise maps and point spread functions (PSFs) modelled by
{\tt TinyTim} \citep{krist93} are those used by the SLI reconstruction
algorithm.

The left-hand column in Figure \ref{recon} shows the resulting F110W
and F160W images for each system. We applied annular elliptical masks
(after PSF convolution), fitted by eye to each image, to include only
the lensed image features. This also masks any noisy residuals which
remain after the {\tt GALFIT} subtraction.

\section{Methodology}
\label{sec_method}

In this paper, we apply the SLI method originally derived by
\citet{warren03}. We use the Bayesian version of the SLI method
applied by \citet{dye07b} and \citet{dye08}, based on the version
developed by \citet{suyu06}.  In addition, the adaptive source plane
grid introduced by \citet{dye05} is used.

In this section, we describe an enhancement to the SLI method that
allows multiple images to be simultaneously reconstructed using the
same lens mass model. Including multiple images in the inversion gives
rise to stronger constraints on the lens model parameters. This is
particularly true if the images are observed at different wavelengths
since colour variations across the lensed source mean that each image
probes a different line of sight through the gravitational
potential of the lensing galaxy. We describe the modifications
necessary for the inclusion of multiple images but refer the reader to
the aforementioned papers for more comprehensive details of the
underlying SLI method.

\subsection{Multi-image SLI method}

The SLI method assumes a pixelised image and a pixelised source plane.
For a given lens model, the method computes the linear superposition
of lensed images of each source plane pixel that best fits the
observed lensed image. In the original formulation, the rectangular
matrix $f_{ij}$ held the fluxes of lensed image pixels $j$ for each
source plane pixel $i$ of unit surface brightness. In this way, a
model lensed image was created with flux values equal to $\sum_i \,
s_i f_{ij}$ for each image pixel $j$ given source pixel surface
brightnesses $s_i$. Subtracting this model image from the observed
image which has pixel flux values $d_j$ and $1\sigma$ uncertainties
$\sigma_j$ allows the $\chi^2$ statistic to be computed.

To cope with multiple images, we need to introduce a new index
to each of these quantities to denote separate image numbers. The
$\chi^2$ statistic in this case becomes
\be
\label{eq_chisq}
\chi^2=\sum_{k=1}^K \left[ \sum_{j=1}^{J_k} \left( 
\frac{\sum_{i=1}^{I_k} s_i^k f_{ij}^k - d_j^k}{\sigma_j^k}\right)^2 \right]
\ee
where there is now an additional sum over images $k$. Note that each
image $k$ has its own source image with surface brightnesses $s_i^k$
in pixels $i$. The image of each of these pixels stored in $f_{ij}^k$
must be convolved with the PSF of the observed image $k$.  Also note
that each of the $k$ sources and $k$ images can have different numbers
of pixels, $I_k$ and $J_k$ respectively.

As in the single image version of the SLI method, the minimum $\chi^2$
solution is given by
\be
\label{eq_recon_simple}
{\rm \mathbf{s}} = {\rm \mathbf{F}}^{-1} {\rm \mathbf{d}}
\ee
but now the matrix ${\rm \mathbf{F}}$ is a block-diagonal matrix
diag$({\rm \mathbf{F}}^1, {\rm \mathbf{F}}^2, ..., {\rm
\mathbf{F}}^K)$ comprising the sub-matrices ${\rm \mathbf{F}}^k$ and
${\rm \mathbf{d}}$ is a column vector which itself has column vector
elements ${\rm \mathbf{d}}^k$. The elements of ${\rm \mathbf{F}}^k$
and ${\rm \mathbf{d}}^k$ are respectively
\be
F_{ij}^{k}=\sum_{n=1}^{J_n}\,f_{in}^k f_{jn}^k / (\sigma_n^k)^2
\, \, , \quad
d_i^k=\sum_{n=1}^{J_n}\,f_{in}^k d_{n}^k / (\sigma_n^k)^2 \, .
\ee
Finally, the column vector ${\rm \mathbf{s}}$ contains the
source pixel surface brightnesses arranged in order $s_1^1, s_2^1, ...
s_{I_1}^1, s_1^2, ... s_{I_K-1}^K, s_{I_K}^K$.

To regularise the solution, equation (\ref{eq_recon_simple}) must be
modified by the regularisation matrix ${\rm \mathbf{H}}$ as described
in \citet{warren03}. However, in the case of multiple images, ${\rm
\mathbf{H}}$ becomes the block diagonal matrix diag$(\lambda_1{\rm
\mathbf{H}}^1, \lambda_2{\rm \mathbf{H}}^2, ..., \lambda_K{\rm
\mathbf{H}}^K)$ where each sub-matrix ${\rm \mathbf{H}}^k$
corresponds to the chosen regularisation matrix for source $k$. Note
that in this formalism, each source $k$ is assigned its own
independent regularisation weight $\lambda_k$. We regularise
each source plane using the scheme described in \citet{dye08}
appropriate for adaptive source grids. In principle,
instead of regularising each source plane independently, different
source planes could be allowed to regularise one another in 
which case ${\rm \mathbf{H}}$ would not be block-diagonal. This
could be beneficial if the source is expected to be similar between
the different input images but would bias the lens model
if not.

The procedure for finding the most probable lens model parameters
then turns to Bayesian inference as described by 
\citet{dye08}. Adapting equation (7) in \citet{dye08} to the
multi-image case here results in the following expression for
the Bayesian evidence, $\epsilon$,
\ba
\label{eq_evidence}
-2 \,{\rm ln} \, \epsilon &=& \chi^2
+{\rm ln} \, \left[ {\rm det} (\mathbf{F}+\mathbf{H})\right]
-{\rm ln} \, \left[ {\rm det} (\mathbf{H})\right]
\nonumber \\
& & 
+ \mathbf{s}^{T}\mathbf{H\,s} + \sum_{k=1}^K\sum_{j=1}^{J_k} 
{\rm ln} \left[2\pi (\sigma_j^k)^2 \right] \, 
\ea
with $\chi^2$ given by equation (\ref{eq_chisq}) and where
${\rm \mathbf{F}}$, ${\rm \mathbf{H}}$ and ${\rm \mathbf{s}}$
are the multi-image quantities defined above. The negative
logarithm of the evidence as given above is
then minimised by applying the Marcov Chain Monte Carlo (MCMC) 
technique to the lens model parameters (see next section),
the regularisation weights\footnote{In practice, we have opted
to set the same regularisation weight across all images to simplify
the MCMC minimisation.} and a parameter called the `splitting
factor' which controls the distribution of source plane pixel
sizes on the adaptive grid \citep[see][for more details]{dye08}.
After the MCMC chain has burnt in, we allow a further 100,000
iterations to estimate parameter confidences.

We note a further practicality. When computing the $\chi^2$ term in
equation (\ref{eq_evidence}), we carry out the sum over image pixels
contained within an annular mask surrounding the ring.  The mask is
tailored for each image to include the image of the entire source
plane, with minimal extraneous sky. This increases the fraction of
significant image pixels with the effect that the evidence is more
sensitive to the model parameters.

Finally, we point out that the multi-image SLI method as presented
assumes that all images are statistically independent of each other.
In the case of images that are not statistically independent, for
example, as could be the case with image slices in an integral field
unit data cube or spectral-line interferometric data cube, equation
(\ref{eq_chisq}) must be modified to include the relevant covariance
terms.

\subsection{Lens model}

We model each of the five lenses considered in this work with a single
smooth density profile to describe the distribution of the total
(baryonic and dark) lens mass. In order to directly compare with the
work of \citet{bolton12} and \citet{sonnenfeld13}, we use the power-law
density profile assumed in these studies. The volume mass density of
this profile, $\rho$, scales with radius, $r$, as $\rho \propto
r^{-\alpha}$. The implicit assumption made with this profile is that
the power-law slope, $\alpha$, is scale invariant. This
assumption appears to be reasonable, at least on the scales probed by
strong lensing, since there is no apparent trend in slope with
the ratio of Einstein radius to effective radius
\citep{koopmans06,ruff11}.

The corresponding projected mass density profile we therefore use in
the lens modelling is the elliptical power-law profile introduced by
\citet{Ka93} which has a surface mass density, $\kappa$, given by
\be
\kappa=\kappa_0\,({\tilde r}/{\rm 1kpc})^{1-\alpha} \, .
\ee
where $\kappa_0$ is the normalisation surface mass density (the
special case of $\alpha=2$ corresponds to the singular isothermal
ellipsoid). The radius ${\tilde r}$ is the elliptical radius defined
by ${\tilde r}^2 =x^{\prime2}+y^{\prime2}/\epsilon^2$ where $\epsilon$
is the lens elongation defined as the ratio of the semi-major to
semi-minor axes. There are three further parameters that describe
this profile: the orientation of the semi-major axis measured in a
counter-clockwise sense from north, $\theta$, and the co-ordinates of
the centre of the lens in the image plane, $(x_c,y_c)$. We also
include two further parameters to allow for an external shear field,
namely, the shear strength, $\gamma$, and shear direction angle, again
measured counter-clockwise from north, $\theta_\gamma$. The shear
direction angle is defined to be perpendicular to the direction of
resulting image stretch.  This brings the total number of lens model
parameters to eight. We assume a uniform prior for all eight
parameters. In the MCMC contour plots presented in the
appendix, we marginalise over $(x_c,y_c)$ since we did not
detect any significant offsets between the lens mass centre and
the centroid of the lens galaxy light.

\section{Results}
\label{sec_results}

The reconstruction of each of the five lenses is shown in Figure
\ref{recon}. Table \ref{tab_results} gives the lens model parameters,
including the geometric average Einstein radius, $\theta_{\rm E}$,
computed as 
\be
\label{eq_r_ein}
\left(\frac{\theta_{\rm E}}{1\,\rm kpc}\right)=
\left(\frac{2}{3-\alpha} \, \frac{1}{\sqrt{\epsilon}} \, 
\frac{\kappa_o}{\Sigma_{\rm CR}}\right)^{\frac{1}{\alpha-1}}
\ee
where $\Sigma_{\rm CR}$ is the critical surface mass density \citep[see, for
example][]{Sc92}.

\begin{table*}
\centering
\small
\begin{tabular}{lcccccccc}
\hline
ID & $z_{\rm d}$ & $z_{\rm s}$ & $\alpha$ & $\kappa_0$ & $\epsilon$ & $\theta (^\circ)$ & $\gamma$ & $\theta_{\rm E}('')$ \\
\hline
J090311.6$+$003906 (ID81)  & 0.2999 & 3.042 & $1.93^{+0.06}_{-0.06}$ & $0.81^{+0.03}_{-0.03}$ & $1.27^{+0.07}_{-0.07}$ & $11^{+8}_{-6}$ & $0.04^{+0.02}_{-0.01}$ & $1.56 \pm 0.11$ \\
J090740.0$-$004200 (ID9)   & 0.6129 & 1.577 & $1.96^{+0.05}_{-0.07}$ & $0.60^{+0.02}_{-0.02}$ & $1.14^{+0.08}_{-0.08}$ & $43^{+3}_{-3}$ & $0.01^{+0.01}_{-0.01}$ & $0.71 \pm 0.05$ \\
J091043.1$-$000321 (ID11)  & 0.7932 & 1.786 & $1.80^{+0.05}_{-0.04}$ & $0.76^{+0.02}_{-0.02}$ & $1.37^{+0.05}_{-0.04}$ & $46^{+3}_{-3}$ & $0.23^{+0.01}_{-0.01}$ & $0.84 \pm 0.04$ \\
J091305.0$-$005343 (ID130) & 0.2201 & 2.626 & $1.74^{+0.20}_{-0.24}$ & $0.26^{+0.02}_{-0.02}$ & $1.34^{+0.18}_{-0.15}$ & $54^{+14}_{-10}$ & $0.02^{+0.02}_{-0.02}$ & $0.43 \pm 0.08$ \\
J090302.9$-$014127 (ID17)  & 0.9435 & 2.305 & $1.37^{+0.21}_{-0.20}$ & $0.31^{+0.04}_{-0.04}$ & $1.29^{+0.15}_{-0.17}$ & $149^{+19}_{-26}$ & $0.03^{+0.03}_{-0.03}$ & $0.36 \pm 0.06$\\
\hline
\end{tabular}
\normalsize
\caption{Lens model parameters. Reading from left to right, columns
  are the H-ATLAS identifier (including the Negrello et al. 2010
  identifier), the lens redshift, $z_{\rm d}$, the source redshift,
  $z_{\rm s}$, the density profile slope, $\alpha$, the lens mass
  normalisation, $\kappa_0$ (in units of
  $10^{10}$M$_\odot$kpc$^{-2}$), the elongation of the lens mass
  profile, $\epsilon$, the orientation of the semi-major axis of the
  lens, $\theta$, measured counter-clockwise from north, the strength
  of the external shear component, $\gamma$, and the Einstein radius,
  $\theta_{\rm E}$, in arcsec computed from equation
  (\ref{eq_r_ein}).}
\label{tab_results}
\end{table*}

\begin{figure*}
\epsfxsize=14cm
{\hfill
\epsfbox{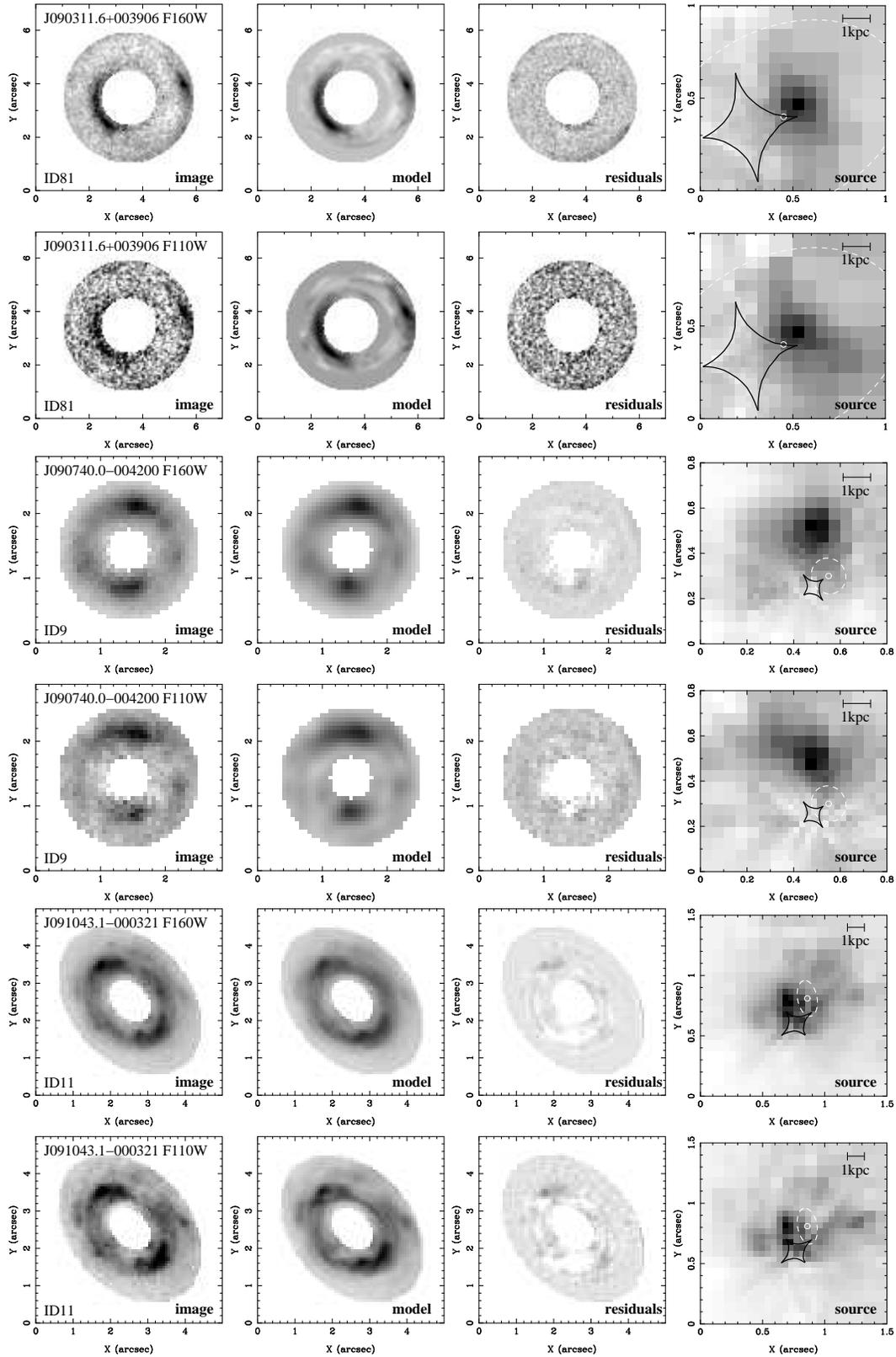}
\hfill}
\epsfverbosetrue
\caption{Lens reconstructions. Reading from left to right, columns
  show the observed image (masked and lens subtracted), the model
  image, the residuals (observed image minus model; grey-scale same as
  corresponding images) and reconstructed source surface brightness
  map (the solid black or white line shows the caustic and the dashed
  white line and small circle respectively show the source half-light
  area and source centre obtained by B13 at 880\,$\mu$m). For each
  system, the F110W and F160W data are shown. In all panels, north
  points along the positive y-axis and west points along the positive
  x-axis.}
\label{recon}
\end{figure*}

\begin{figure*}
\epsfxsize=14cm
{\hfill
\epsfbox{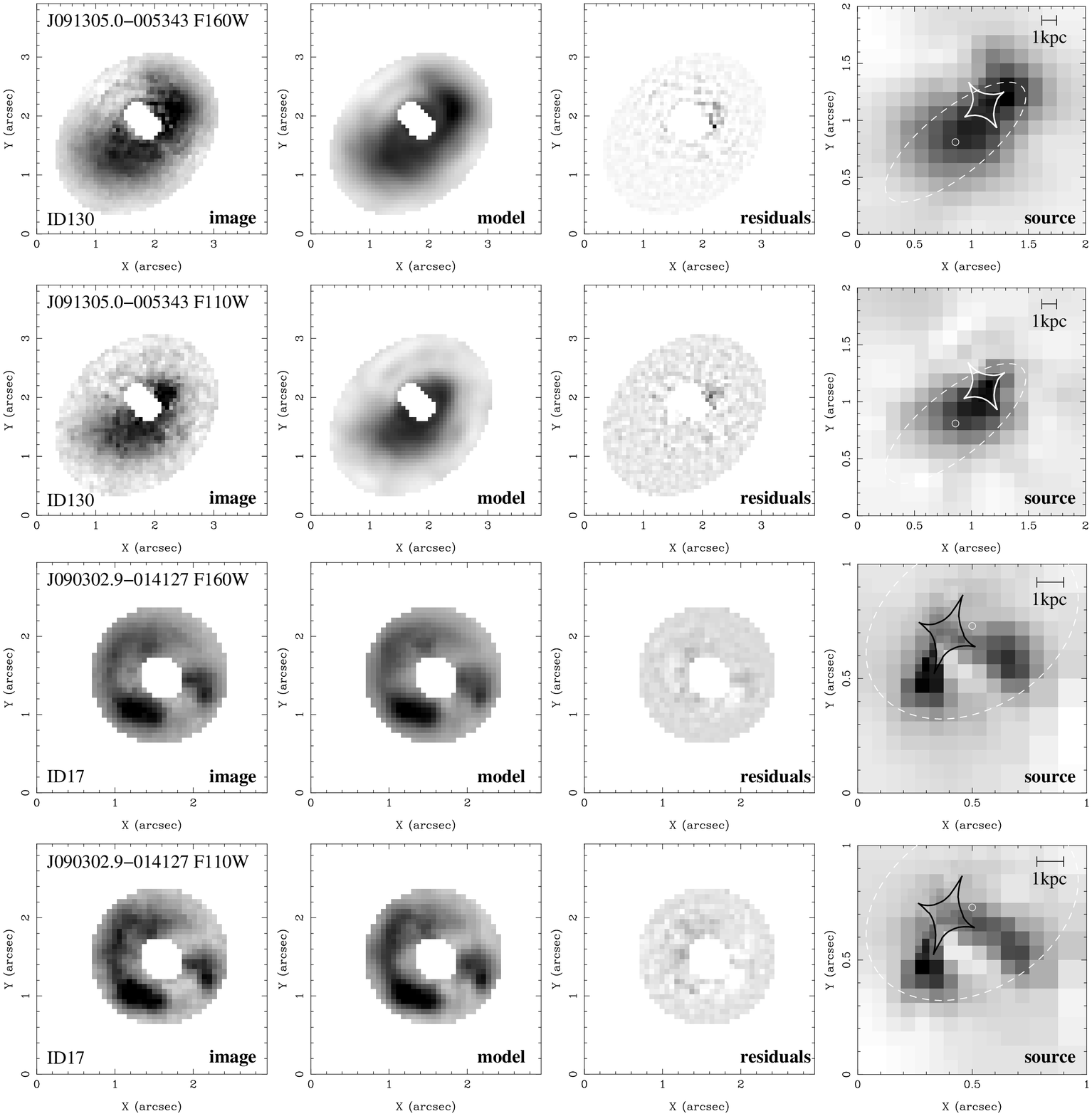}
\hfill}
\epsfverbosetrue
\contcaption{}
\end{figure*}

\subsection{Object Notes}

In this section we detail the characteristics of each lens system.  In
particular, we compare with the results of \citet[][B13
  hereafter]{bussmann13} who have modelled imaging data acquired with
the Sub-Millimeter Array (SMA) for $\sim 30$ candidate lenses
discovered by H-ATLAS and HerMES. All five of the H-ATLAS SDP lenses
are modelled by B13, although we point out that external shear is
not included in their lens model.

{\em J090311.6$+$003906} (ID81): This is a classic cusp-caustic
configuration lens. The near-IR emission in the lensed image closely
matches that of the SMA data and unsurprisingly results in a
consistent magnification. The reconstructed source in both HST filters
shows little structure other than a slight elongation along the NE-SW
direction. The centroid of the source is well aligned with that found
by B13.  The lens model is an excellent fit to the observed data and
leaves no significant residuals. The model requires a small amount of
external shear with strength $\gamma \simeq 0.05$ and direction
$\theta_\gamma \simeq 105^\circ $, consistent with perturbations
expected from a nearby group of galaxies to the east.  The elongation
and orientation of the required model is entirely consistent with that
of the observed light profile.

{\em J090740.0$-$004200} (ID9): This lens system is dominated by a
single doubly-imaged source with a simple morphology lying to the
north of the lens galaxy centroid. The double imaging is consistent
with the SMA data but the emission appears to originate from a
different location in the source compared to what is observed in the
near-IR HST data. The reconstructed source F110W-F160W colour map
shows a reddening gradient which points from the near-IR source
towards the SMA source centroid.  Some of the fainter emission from
the source in the near-IR crosses the caustic and gives rise to the
observed complete Einstein ring. There is also some fainter structure
in the ring which is fit in the model by a fainter source to the east
of the lens centroid. Negligible external shear is required in the
best fit lens model. The modest elongation of $\epsilon=1.14$
indicates a more radially symmetric mass profile compared to the
light. The alignment of the mass and light elongation in this lens is
significantly different (see section \ref{sec_morph}).

{\em J091043.1$-$000321} (ID11): Like ID9, this system has a complete
Einstein ring. The ring's significant ellipticity is the result of a
relatively strong external shear field of strength $\gamma=0.23$. The
direction of this shear points almost exactly to the centre of a
nearby edge-on spiral galaxy to the NW located at a redshift of
$z=0.39\pm0.09$ (see N13). The implication is therefore that this
spiral is almost entirely responsible for the shear perturbation. We
attempted a model where external shear was replaced by a second
singular isothermal lens to represent the spiral's total mass but
found no significant improvement in the fit. The reconstructed source
exhibits clear small scale structure in both filters, required to fit
the observed structure in the ring. The majority of the emission
observed in the ring comes from a doubly imaged source lying just
outside the caustic and the smaller-scale structure comes from smaller
knots of emission in the source, some of which are quadruply
imaged. As the residual plot in Figure \ref{recon} indicates, the
model image does not perfectly account for the observed features in
the ring and this is also reflected in the fact that the model fit is
only marginally acceptable. It is therefore possible that some of the
ring structure is actually structure in the lens galaxy not fully
removed by {\tt GALFIT}.  The SMA data imply a doubly imaged source
which, like J090740.0$-$004200, is offset from the near-IR emission
but has a comparable Einstein radius. As with ID9, the reconstructed
source F110W-F160W colour map shows a reddening gradient which points
from the near-IR source towards the SMA source centroid.

In addition to the reconstructed near-IR source presented here, there
appears to be further lensed arcs associated with this system (not
shown in Figure \ref{recon}). Two readily identified arcs lie to the
south and to the east of the lens centre and trace out a ring with an
Einstein radius larger than that shown in Figure \ref{recon}. This
implies that the lensed source responsible lies at a different
redshift. We chose not to complicate the lens model further by
incorporating this additional source but instead leave this for
further work.

{\em J091305.0$-$005343} (ID130): The lens galaxy is relatively poorly
constrained in this system. The best fit model is consistent with zero
external shear (as might be expected from the lack of observed nearby
perturbers) and the lens has one of the higher elongations found in
the sample of $\epsilon = 1.34$. The best fit reconstructed source is
very extended and as such, the magnification is low. The residual map
shown in Figure \ref{recon} shows some significant features towards
the lens centre which is not surprising given the difficulty reported
by N13 in removing the lens light. However, the residuals contribute
an insignificant amount of light to the overall source and as such
will have a negligible effect on the lens model parameters. The SMA
data are in close agreement with the near-IR data which increases
confidence in the {\tt GALFIT} subtraction (see N13), although the
near-IR data imply a slightly larger magnification than the SMA
data. We make the caveat that this lens, as pointed out by N13, is a
likely Sa galaxy.  In order to compare our results with those of the
aforementioned lensing studies which include only early type lenses,
we have omitted this system from our fits to the trends
reported later in this paper.

{\em J090302.9$-$014127} (ID17): This is a relatively poorly
constrained lens although the fit is acceptable. There are some minor
features in the residual image which occur at 3 o'clock and 11 o'clock
around the inner radius of the annulus as shown in Figure
\ref{recon}. N13 have modelled the flux in the lensed image by fitting
individual {\tt GALFIT} profiles to the different components. Two of
these profiles lie on the inner radius of the annulus and these
coincide with the locations of the residuals found in the lens
modelling here. One interpretation is therefore that spatially
dependent extinction in the lensing galaxy affects the lensed
image. This is consistent with the colour of the feature at 11 o'clock
which N13 determine as having a significantly redder colour than the
average colour of the other {\tt GALFIT} profiles, but not with the
feature at 3 o'clock which is consistent in colour. An alternative
explanation might therefore be that the lensed image contains residual
flux from the lensing galaxy itself which is highly possible given the
complexity in removing the lens from this system. Nevertheless, the
reconstructed source plane shows two very prominent elongated objects
in both bands which converge at a point interior to the caustic. The
majority of this source plane emission is doubly imaged but some of
the flux in the merged region is quadruply imaged. The fact that the
lensed image is well fit by a relatively simple source surface
brightness map adds reassurance that the reconstruction is plausible;
an over-complicated source often indicates that there are non-lensed
features in the lensed image. Although the SMA data for this lens are
unable to resolve individual ring features, B13 obtain a magnification
consistent with that measured from the near-IR data here.

\subsection{Mass profile vs. redshift}

\begin{figure}
\epsfxsize=8.5cm
{\hfill
\epsfbox{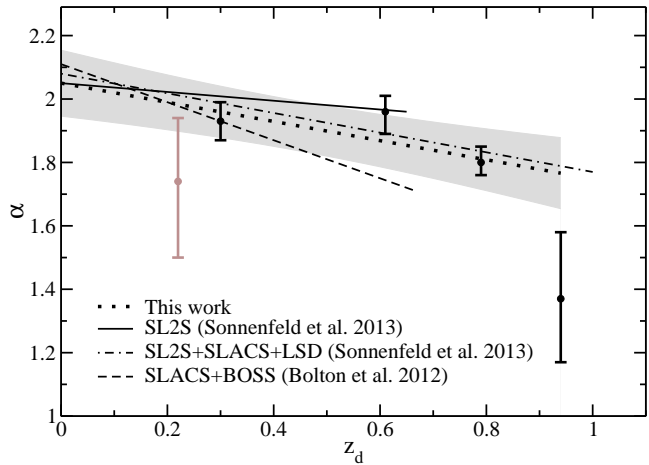}
\hfill}
\epsfverbosetrue
\caption{Variation of the density profile slope, $\alpha$, as a
  function of the lens redshift, $z_{\rm d}$. The solid and dashed
  lines show the redshift dependency of $\alpha$ using the SL2S lenses
  by \citet{sonnenfeld13} and the SLACS $+$ BELLS lenses by
  \citet{bolton12} respectively. The extent of each line indicates the
  extent of the lens redshifts in the respective surveys. The grey
  data point corresponds to the Sa lens J091305.0$-$005343 which has
  been excluded in our fit shown by the dotted line and grey shaded
  1$\sigma$ error envelope.}
\label{slope_evol}
\end{figure}

Figure \ref{slope_evol} shows the fitted lens density profile slopes
plotted against redshift. Discounting the likely Sa lens
J091305.0$-$005343, the straight line minimum $\chi^2$ fit through the
four data points is $\alpha=2.05\pm0.08 - (0.30\pm0.13)z$. (If we
include this fifth lens, the fit becomes $\alpha=2.01\pm0.10 -
(0.25\pm0.15)z$.)  The dotted line and grey shaded envelope in Figure
\ref{slope_evol} shows the fit and the $1\sigma$ error region
respectively. In the same figure we plot the variation in slope with
redshift from three other lens sample combinations; 1) the SL2S lens
sample of $\alpha=2.05\pm0.06-(0.13\pm0.24)z$ from
\citet{sonnenfeld13}, 2) the combination of the Lensing Structure and
Dynamics (LSD) sample of Treu \& Koopmans (2004, ApJ, 611, 739), SL2S
and SLACS of $\alpha=2.08\pm0.02-(0.31\pm0.10)z$ also from
\citet{sonnenfeld13} and 3) the combination of SLACS and BOSS of
$\alpha=2.11\pm0.02-(0.60\pm0.15)z$ from Bolton et al. (2012).

As is apparent from Figure \ref{slope_evol}, our inferred rate of
change in slope with redshift is not inconsistent with that measured
by any of the other studies plotted in the figure, although neither is
it inconsistent with a null rate of change. This is perhaps not
surprising given the small sample size presently at our disposal.
This limitation will be considerably reduced by our forthcoming
rapidly growing lens sample.

\subsection{Lens Magnifications}

Table \ref{tab_mags} lists the source flux magnifications.
For each lens, we have computed the magnification at every point
in the MCMC chain as presented in Figure \ref{params} to form
a magnification distribution. Table \ref{tab_mags} then
quotes the median magnification and its $\pm 34\%$ bounds.

\begin{table*}
\centering
\small
\begin{tabular}{lcccc}
\hline
ID & $\mu_{\rm tot}$ & $\mu_{0.5}$ & $\mu_{0.1}$ & $\mu_{880}$ \\
\hline
J090311.6$+$003906 (ID81)  & $10.6^{+0.6}_{-0.7}$ & $13.8^{+1.0}_{-0.9}$ 
      & $21.0^{+1.6}_{-1.4}$ & $11.1 \pm 1.1$ \\
J090740.0$-$004200 (ID9)   & $6.29^{+0.27}_{-0.26}$ & $7.23^{+0.31}_{-0.34}$  
      & $5.80^{+0.38}_{-0.27}$ & $8.8 \pm 2.2$ \\
J091043.1$-$000321 (ID11)  & $7.89^{+0.21}_{-0.25}$ & $12.5^{+0.40}_{-0.43}$ 
      & $28.3^{+2.1}_{-3.5}$ & $10.9 \pm 1.3$ \\
J091305.0$-$005343 (ID130) & $3.09^{+0.22}_{-0.21}$  & $4.01^{+0.25}_{-0.25}$ 
      & $7.51^{+0.32}_{-0.37}$ & $2.1 \pm 0.3$ \\
J090302.9$-$014127 (ID17)  & $3.56^{+0.19}_{-0.17}$ & $4.48^{+0.33}_{-0.25}$ 
      & $5.54^{+0.41}_{-0.30}$ & $4.9 \pm 0.7$ \\
\hline
\end{tabular}
\normalsize
\caption{Source flux magnifications. The quantities listed are:
  the total source flux magnification, $\mu_{\rm tot}$; the
  magnifications, $\mu_{0.5}$ and $\mu_{0.1}$ which give the
  magnification of the brightest region(s) of the source respectively
  contributing 50\% and 10\% of the total reconstructed source flux;
  the magnification at 880$\,\mu$m, $\mu_{880}$,
  computed by B13 for an area of the source which is four
  times the source's half-light area.}
\label{tab_mags}
\end{table*}

We computed magnifications using the higher signal-to-noise F160W data
to give more precise magnifications. We find that the magnifications
computed using the F110W data generally have a larger spread but the
distribution is always consistent with those computed using the F160W
data.

We determined three different magnifications to give an indication of
the strength of near-IR differential amplification. The first is a
`total magnification', $\mu_{\rm tot}$, computed as the ratio of the
total flux in the masked region of the image as shown in Figure
\ref{recon} to the total flux in the source plane. The second and
third, $\mu_{0.5}$ and $\mu_{0.1}$, correspond to the magnification of
the brightest region(s) of the source that contributes 50\% and 10\%
respectively of the total reconstructed source flux.  Note that
$\mu_{\rm tot}$ is almost always lower than $\mu_{0.5}$ and
$\mu_{0.1}$ since incorporating the total source plane typically
includes additional regions that are less magnified.

Table \ref{tab_mags} also lists the magnifications, $\mu_{880}$,
determined by B13 at 880\,$\mu$m.  These magnifications are calculated
in an elliptical disk centred on the best fit 880\,$\mu$m source
brightness profile with a radius twice that containing half of the
total source flux. $\mu_{880}$ is the ratio of the integrated flux
within the image plane region mapped by this disk to the integrated
flux within the disk in the source plane itself.  For the S\'{e}rsic
profiles fit to the SDP sources by B13, this corresponds to
approximately 75\% of the source light in all cases. Therefore,
$\mu_{880}$ corresponds to a magnification somewhere between $\mu_{\rm
  tot}$ and $\mu_{0.5}$.

As Table \ref{tab_mags} shows, for all lenses apart from
J091305.0$-$005343, $\mu_{880}$ is consistent with a value spanned by
$\mu_{\rm tot}$ and $\mu_{0.5}$. The consistency is generally better
when the 880$\,\mu$m source morphology more closely resembles the
reconstructed near-IR source. This is a reflection of the fact that
the lens models determined at both wavelengths are geometrically
similar\footnote{We note that the lens elongations derived by B13 are
consistently higher than those derived in our study and we attribute
this, at least partly, to the lack of external shear in the B13 lens
model.}.  The exception is J091305.0$-$005343 which has a very
similar source morphology between 880$\,\mu$m and the near-IR, but in
this case, the magnification discrepancy is brought about by a
significantly different lens elongation, with the near-IR lens model
favouring an elongation of $\sim 1.3$ compared to $\sim 2.0$ at
880$\,\mu$m.

\subsection{Comparison of mass and light morphology}
\label{sec_morph}

A question which brings insight to models of galaxy formation and
evolution is how closely the visible mass traces the dark matter
halo. One way to address this is to compare the visible morphology
of the lens galaxies to the total mass profiles determined through
lensing.

We used the {\tt GALFIT} models of N13 to determine the elongation and
orientation of the lens galaxy surface brightness profiles. Table
\ref{tab_light_morph} lists these along with the effective radii. In
the case of J091305.0$-$005343 (ID130), we used only the light
profiles which make up the bulge, since the bulge comprises nearly all
of the light and, unlike the faint disk, has a coherent set of
profiles which give a well-defined orientation and elongation.

\begin{table*}
\centering
\small
\begin{tabular}{lccccccc}
\hline
ID & $\epsilon_{\rm light}$ & $\theta_{\rm light} (^\circ)$ & 
     R$_{\rm e}$ (``) & R$_{\rm e}$ (kpc) & M$^{\rm *,Salp}_{\rm R_e/2}$ & 
     M$^{\rm *,Chab}_{\rm R_e/2}$ & M$^{\rm Tot}_{\rm R_e/2}$ \\
\hline
J090311.6$+$003906 (ID81) & $1.24 \pm 0.01$ & $10 \pm 1$ & 0.45 & 2.1 & $4.36\pm1.30$ & $2.45\pm0.73$ & $6.46\pm0.24$ \\
J090740.0$-$004200 (ID9)  & $1.31 \pm 0.01$ & $59 \pm 1$ & 0.41 & 2.9 & $3.56\pm1.07$ & $2.00\pm0.60$ & $6.17\pm0.21$ \\
J091043.1$-$000321 (ID11) & $1.58 \pm 0.01$ & $21 \pm 2$ & 0.38 & 3.0 & $4.08\pm2.92$ & $2.29\pm1.64$ & $8.91\pm0.23$ \\
J091305.0$-$005343 (ID130)& $1.18 \pm 0.02$ & $43 \pm 4$ & 0.31 & 1.2 & $2.14\pm0.62$ & $1.20\pm0.35$ & $1.26\pm0.10$ \\
J090302.9$-$014127 (ID17) & $1.79 \pm 0.01$ & $16 \pm 2$ & 0.40 & 3.2 & $1.55\pm0.59$ & $0.87\pm0.33$ & $3.31\pm0.42$ \\
\hline
\end{tabular}
\normalsize
\caption{Morphology of the lens light profile and lens masses. Columns
  are: elongation, $\epsilon_{\rm light}$, position angle (measured
  counter-clockwise from north), $\theta_{\rm light}$, effective
  radius, R$_{\rm e}$, the stellar mass contained within a radius of
  R$_{\rm e}$/2 for a Salpeter and Chabrier IMF, M$^{\rm *,Salp}_{\rm
    R_e/2}$ and M$^{\rm *,Chab}_{\rm R_e/2}$ respectively (derived
  from the stellar masses computed in N13) and the total mass within a
  radius of R$_{\rm e}$/2 inferred from the lens model, M$^{\rm
    Tot}_{\rm R_e/2}$. All masses are in units of $10^{10}{\rm
    M}_\odot$. Note that the light profile parameters for
  J091305.0$-$005343 refer only to the bulge component and exclude the
  faint disk.}
\label{tab_light_morph}
\end{table*}

\begin{figure}
\epsfxsize=8.5cm
{\hfill
\epsfbox{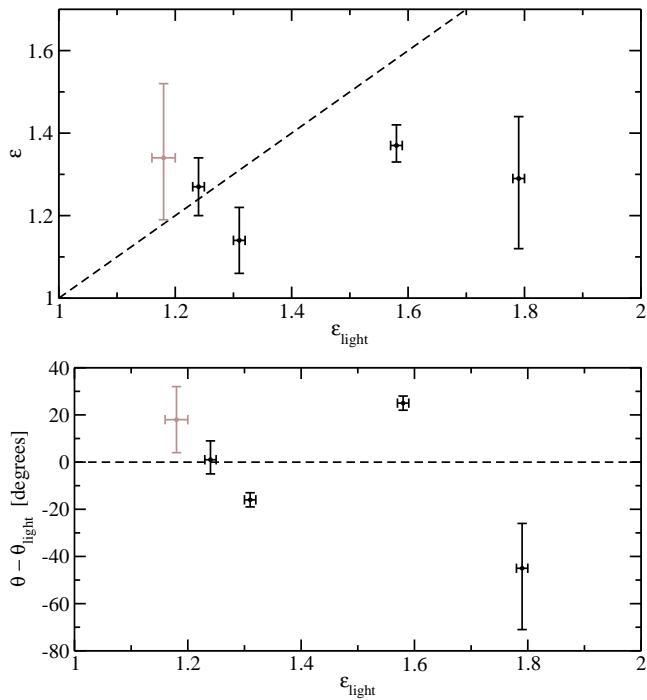}
\hfill}
\epsfverbosetrue
\vspace{-5mm}
\caption{Comparison of light with mass. {\em Top panel}: Elongation of
  lens mass profile versus elongation of observed light.
  {\em Bottom panel}: Difference in
  position angle of lens profile and observed light versus elongation
  of observed light. In both panels, the grey data point corresponds
  to the Sa lens J091305.0$-$005343.}
\label{compare_morph}
\end{figure}

Figure \ref{compare_morph} plots the comparison of mass and light
profile parameters. The top panel shows the comparison of elongations.
For all five of the lenses, the total lens mass model has or is
consistent with a lower elongation than that of the light. This
implies that the dark matter halo is comparable in elongation
or rounder in each case. 

The bottom panel of Figure \ref{compare_morph} compares the offset in
orientation of the lens mass and light profiles. Here, there are some
significant discrepancies. The offsets are substantially higher on
average than those found by SLACS, who measured an rms scatter of
10$^\circ$ \citep{koopmans06}, but consistent with the findings of
SL2S \citep{gavazzi12} who measure an rms scatter of 25$^\circ$ with
offsets of up to 50$^\circ$. As \citet{gavazzi12} point out, the SL2S
lenses have a higher average ratio of Einstein radius to effective
radius than the SLACS lenses and hence the SLACS lenses are more
dominated by the stellar component. The average of this ratio for our
lenses is even higher than that of the SL2S sample and so we would
expect even less alignment between the dark and visible components. We
will be able to explore this trend more properly with our forthcoming
larger lens sample, although there are already indications from
simulations that such large (and even larger) morphological
differences between baryons and dark matter are commonplace \citep[for
example, see][]{bett10, skibba11}.

\subsection{Other observational trends}

An important observational benchmark for models of galaxy and
structure formation is the fraction of dark matter contained within a
fixed fraction of the effective radius. Using the SLACS lens sample,
\citet{auger10b} measure an average projected fraction of dark matter
within half the effective radius of 0.21 with a scatter of 0.20, for a
\citet{salpeter55} IMF, or 0.55 with a scatter of 0.11 for a
\citet{chabrier03} IMF. \citet{ruff11} measure the average of this
fraction to be 0.42 with a scatter of 0.20 for a Salpeter IMF for the
SL2S lenses.

Table \ref{tab_light_morph} lists the stellar masses and the total
lensing projected mass contained within half the effective radius for
our lenses. To obtain these stellar masses, we used the total stellar
masses and the {\tt GALFIT} profiles determined for the lens galaxies
by N13. Excluding the Sa lens J091305.0$-$005343, our lenses have a
mean projected dark matter fraction within half the effective radius
of $f_{\rm DM}=0.46$ with a scatter of 0.10, for a Salpeter IMF,
($f_{\rm DM}=0.69$ with a scatter of 0.07, for a Chabrier IMF).
Although this is statistically consistent with the SLACS and SL2S
lenses, the higher value measured in both our lenses and the SL2S
lenses compared to the SLACS sample most likely reflects the lower
average ratio of Einstein radius to effective radius in SLACS.

One of the trends detected by the SLACS survey is that $f_{\rm DM}$
for early types increases with galaxy mass and galaxy size.
\citet{auger10b} measure the linear fit $f_{\rm DM}=-0.13\pm0.09 +
(0.49\pm0.10)\log({\rm R_e/1\,kpc})$. However, this result is not
confirmed by the SL2S lens sample; \citet{ruff11} fail to find any
correlation between $f_{\rm DM}$ and ${\rm R_e}$ with a measured
gradient of ${\rm d}f_{\rm DM}/{\rm
  d\,R_e}=0.08^{+0.10}_{-0.08}$\,kpc$^{-1}$.  Our lens sample,
excluding the Sa lens J091305.0$-$005343 gives a linear fit of $f_{\rm
  DM}=-0.01\pm0.20 + (1.04\pm0.46)\log({\rm R_e/1\,kpc})$, steeper but
consistent with the SLACS lenses.

\begin{figure}
\epsfxsize=8.5cm
{\hfill
\epsfbox{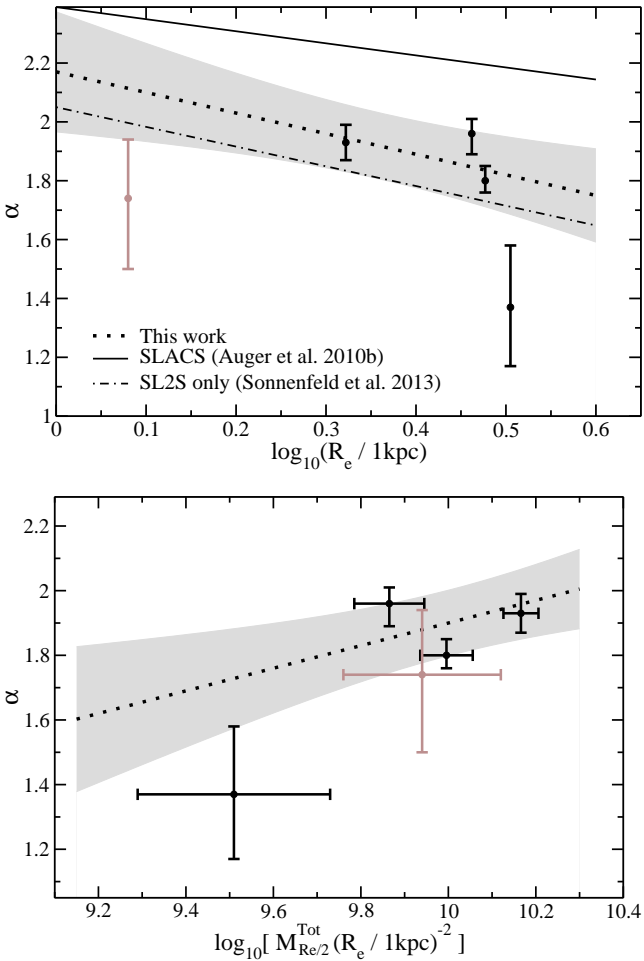}
\hfill}
\epsfverbosetrue
\vspace{-5mm}
\caption{Correlation between total mass density profile slope,
  $\alpha$, and effective radius, R$_{\rm e}$, ({\em top panel}) and
  average projected total surface mass density within R$_{\rm e}/2$
  ({\em bottom panel}). In both panels, the Sa lens
  J091305.0$-$005343, coloured with a grey data point, has been
  omitted in the straight line fit which is shown by the dotted line
  and grey shaded 1$\sigma$ error envelope.}
\label{slope_corr}
\end{figure}

The SLACS and SL2S lenses also exhibit the trend that the slope of the
density profile inferred from lensing is negatively correlated with
effective radius and positively correlated with the average surface
mass density contained within ${\rm R_e}/2$. In Figure
\ref{slope_corr}, we plot these correlations for our lenses. The top
panel shows the density profile slope, $\alpha$, plotted against
effective radius. For our four early-type lenses, we obtain a straight
line fit of $\alpha=2.17\pm0.20 - (0.70\pm0.47)\log({\rm R_e /
  1\,kpc})$, as shown in the figure by the dotted line and grey shaded
$1\sigma$ error envelope.  This compares to the fit
$\alpha=2.39\pm0.10 - (0.41\pm0.12)\log({\rm R_e / 1\,kpc})$ by
\citet{auger10b} for the SLACS lenses and $\alpha=2.05\pm0.06 -
(0.67\pm0.20)\log({\rm R_e / 1\,kpc})$ by \citet{sonnenfeld13} for the
SL2S lenses\footnote{Here, we have taken a slice through the
4-dimensional plane that Sonnenfeld et al. fit to the density profile
slope by assuming a redshift of 0.3 and a stellar mass of $\log({\rm
M_*/M_\odot})=11.5$.}.

In the bottom panel of Figure \ref{slope_corr}, we plot $\alpha$
against the average total surface mass density within half the
effective radius, as quantified by the ratio
M${\rm^{Tot}_{R_e/2}/(R_e/1\,kpc)^2}$.  We obtain a straight line fit
of $\alpha=11.60\pm0.11+ (0.35\pm0.22)\log({\rm
  M^{Tot}_{R_e/2}/(R_e/1\,kpc)^2})$ which compares to the gradient of
${\rm d \alpha / d \log[M^{Tot}_{R_e/2}/(R_e/1\,kpc)^2]=0.85\pm0.19}$
for the SLACS lenses as measured by \citet{auger10b}.

\section{Discussion}

As the preceding section has shown, in all observational diagnostics
and trends we have considered, bar the correlation between $f_{\rm
DM}$ and ${\rm R_e}$, the H-ATLAS lenses are more
similar to the SL2S lenses than those of SLACS. This is perhaps not
surprising when one considers the following characteristic median values
expressed in order SLACS, SL2S, H-ATLAS: 
$\widetilde{\rm R}_{\rm e} \simeq$\,8\,kpc, 5\,kpc, 3\,kpc; 
$\widetilde{\rm M}_* \simeq 10^{11.6}\,{\rm M_\odot}$, 
     $10^{11.5}\,{\rm M_\odot}$, $10^{11.2}\,{\rm M_\odot}$;
$\widetilde{\rm M}^{\rm Tot}_{\rm R_e/2} \simeq 10^{11.2}\,{\rm M_\odot}$, 
     $10^{11.0}\,{\rm M_\odot}$, $10^{10.8}\,{\rm M_\odot}$.
It appears to be the case therefore, at least in the SDP data, that
the H-ATLAS lenses populate the low-mass tail of the SL2S lens
sample.

Despite these obvious differences and despite our very small sample of
lenses at present, we still detect many of the correlations found in
the various other aforementioned studies. These studies have taken
care to ensure that the trends they detect are not the result of
selection biases or systematic effects. In a similar vein, a potential
systematic effect to be considered when comparing our results with
these is that our lensing analysis does not incorporate any additional
constraints from dynamical measurements. This means that the slope is
measured in the vicinity of the Einstein ring, whereas in analyses
using lensing and dynamics, the average slope interior to the Einstein
ring is measured. Therefore, a change in slope with radius could
potentially introduce a systematic offset in the slopes determined in
the present work with respect to those from lensing and
dynamics. However, as previously mentioned, on the scales probed by
strong lensing, the slope appears not to exhibit any significant
dependency on radius since there is no apparent trend in slope with
the ratio of Einstein radius to effective radius
\citep{koopmans06,ruff11}.

This is related to the effect reported in \citep{ruff11} and
\citep{bolton12} that the ratio of the Einstein radius to the lens
galaxy's effective radius increases with increasing lens redshift.
This is due to the redshift dependence of the angular diameter
distance ratios which govern the lensing geometry and the fact that a
fixed physical size reduces in angular extent with increasing redshift
(at least out to the lens redshifts in this work). This has the result
that as redshift is increased, the density profile is measured by our
lensing-only analysis at a radius which is an increasing multiple of
the effective radius. A change in slope with radius would therefore
mimic a change in slope with redshift.  However, in addition to the
observational evidence that the slope is not seen to depend on radius
on strong lensing scales, on much larger scales, the slope is expected
to steepen with increasing radius according to simulations and the
requirement that the total halo mass converges. Therefore, even if
this steepening were to influence our slope measurements, our
detection of the rate at which slopes become less steep with
increasing redshift must be a lower limit to the intrinsic rate.

In terms of a physical interpretation of observed variations in the
density profile slope, the picture is somewhat unclear.  Simulations by
\citet{dubois13} reproduce the observed steepening with decreasing
redshift and find that feedback from active galactic nuclei (AGN)
modifies the slope. This work indicates that AGN feedback is required
to reproduce the near-isothermal profiles (i.e., $\alpha \simeq 2$)
observed in low redshift early type galaxies. However, the simulations
of \citet{remus13} indicate that whilst a combination of dry minor and
major mergers produce near-isothermality at low redshifts, the slopes
are significantly steeper at higher redshift. Confounding this is the
simulation work of \citet{nipoti09} which shows that the total mass
profile of early types is not modified at all by dry mergers.

Turning to the projected dark matter fraction within half the
effective radius, $f_{\rm DM}$, \citet{dubois13} claim that AGN
feedback is required to reproduce the observed fractions and that
without it, the fraction of stellar mass is too high.  The SLACS work
reports a $5\sigma$ detection of increasing $f_{\rm DM}$ with
effective radius.  This compares to our marginal detection
($2.3\sigma$) and a null detection in the SL2S lenses. If this trend
is real, an obvious interpretation might be that star formation
efficiency reduces as halo mass increases. Another possibility is
presented by \citet{nipoti09} who predict that the fraction of dark
matter within the effective radius increases as a result of mergers.

\begin{figure}
\epsfxsize=8.5cm
{\hfill
\epsfbox{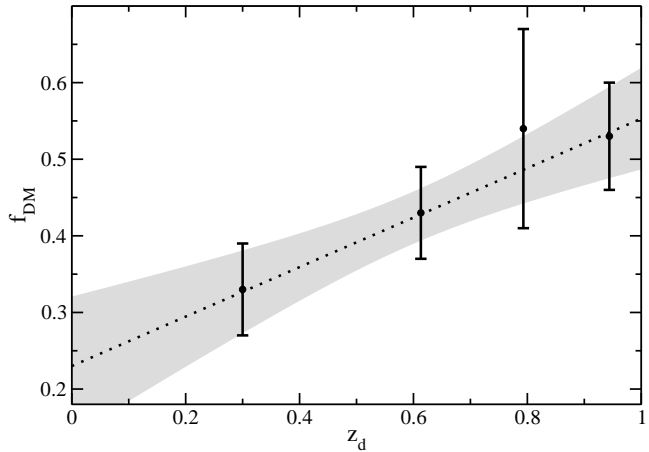}
\hfill}
\epsfverbosetrue
\vspace{-3mm}
\caption{Variation of the fraction of dark matter within half
the effective radius, $f_{\rm DM}$, for a Salpeter IMF with redshift.
The grey shading depicts the $1\sigma$ error region for the straight
line fit.}
\label{fdm_vs_z}
\end{figure}

Instead of investigating how $f_{\rm DM}$ varies with effective
radius, an alternative is to test whether $f_{\rm DM}$ changes with
redshift since this is another diagnostic which can be provided by
simulations.  

Figure \ref{fdm_vs_z} shows this plot for the H-ATLAS lenses
(excluding J091305.0$-$005343). We measure a straight line fit of
$f_{\rm DM}=0.23\pm0.09 + (0.32\pm0.14)z$. In comparison, the
simulations of \citet{dubois13} predict that the fraction of dark
matter within 10\% of the virial radius decreases with increasing
redshift when AGN feedback is present, or, that this fraction remains
constant with redshift if AGN feedback is not present. If the fraction
of dark matter within 10\% of the virial radius scales in the same way
as $f_{\rm DM}$, then this is in contrast to our findings.  However,
since $f_{\rm DM}$ depends on the size of the stellar component and
the virial radius effectively does not, there is still the possibility
that the two results are consistent if the stellar mass increases in
spatial extent relative to the dark matter with increasing redshift.

Important clues also come from comparing the morphology of the visible
component of the lenses with that of the dark matter halo.  We find
significant discrepancies in the alignment and ellipticity between the
stellar component and the total mass in some lenses. The discrepancies
are consistent with what has been measured in the SL2S sample but
larger than those found in SLACS. This may be a combination of the
fact that both the H-ATLAS and SL2S lenses have a higher average ratio
of Einstein radius to effective radius than the SLACS lenses and that
the baryonic morphology correlates less strongly with that of the dark
matter at larger radii \citep[e.g.][]{bett10, skibba11}.  In order to
proceed with a more robust interpretation of these findings, more
input is required from simulation work although as we previously
discussed, present indications are that such large morphological
differences between the dark and baryonic components are to be
expected.

\section{Summary}
\label{sec_summary}

In this paper, we have modelled the first five strong gravitational
lens systems discovered in the H-ATLAS SDP data. To directly compare
with other lensing studies, we have modelled the lenses with
elliptical power-law density profiles and searched for trends in the
power-law slope and the fraction of dark matter contained within half
the effective radius. We have found consistency with almost all
existing lens analyses, although with our present sample of only five
lenses, we lack high statistical significance in our measured
trends. The main results of this paper are that:

\begin{itemize}

\item the slope of the power-law density profile varies with redshift
  according to $\alpha=2.05\pm0.08 - (0.30\pm0.14)z$.

\item the H-ATLAS lenses have a mean projected dark matter fraction
  within half the effective radius of $f_{\rm DM}=0.46$ with a scatter
  of 0.10, for a Salpeter IMF.

\item the dark matter fraction within half the effective radius scales
  with effective radius as $f_{\rm DM}=-0.01\pm0.20 +
  (1.04\pm0.46)\log({\rm R_e/1\,kpc})$.

\item the slope of the power-law density profile scales with effective
  radius as $\alpha=2.17\pm0.20 - (0.70\pm0.47)\log({\rm R_e /
    1\,kpc})$ and with the average total surface mass density within
  half the effective radius, as quantified by the ratio
  M${\rm^{Tot}_{R_e/2}/(R_e/1\,kpc)^2}$ as
  M${\rm^{Tot}_{R_e/2}/(R_e/1\,kpc)^2}$.

\item $f_{\rm DM}$ scales with redshift as $f_{\rm DM}=0.23\pm0.09 +
  (0.32\pm0.14)z$.

\end{itemize}

The modelling in this paper used near-IR HST data. Whilst the HST
provides the high resolution imaging necessary for modelling of high
redshift lenses, not all of the H-ATLAS lensed sources will be as
readily detected in the near-IR as the SDP lenses considered herein.
Being submm selected systems, submm and radio interferometry is
the ideal technology for obtaining the required signal to noise and
image resolution. This has been demonstrated by \citet{bussmann13} who
have used the SMA to image several tens of lenses detected by the
Herschel Space Observatory. ALMA has also been used to image some of
the SPT lenses \citep[see, for example][]{hezaveh13}. However, the
true power of this facility will not be realised until it operates
with its full complement of antennae. At this point, ALMA will begin
to deliver the high signal-to-noise and high resolution images
required by source-inversion lens modelling methods, necessary for the
strongest possible constraints on galaxy mass profiles. Furthermore,
spectral line imaging with ALMA will open up the possibility of
reconstructing lensed source velocity maps to probe the dynamics of
high redshift submm galaxies.

The H-ATLAS lens sample is very much in its infancy. As the size of
the sample grows and begins to populate the $z_{\rm d} \simeq 1$ realm
and beyond, constraints on the evolution of mass in galaxies will
continue to strengthen.

\appendix
\section{Lens parameter confidence plots}
\label{app_lens_pars}

In this appendix, we plot the confidence contours for all parameter
combinations for each lens (apart from the lens position parameters
$x_c$ and $y_c$ since we did not detect any significant offsets
between the lens mass centre and the centroid of the lens galaxy
light). In each plot, the contours correspond to the 1, 2 and
3$\sigma$ confidence levels.

\begin{figure*}
\epsfxsize=17.8cm
{\hfill
\epsfbox{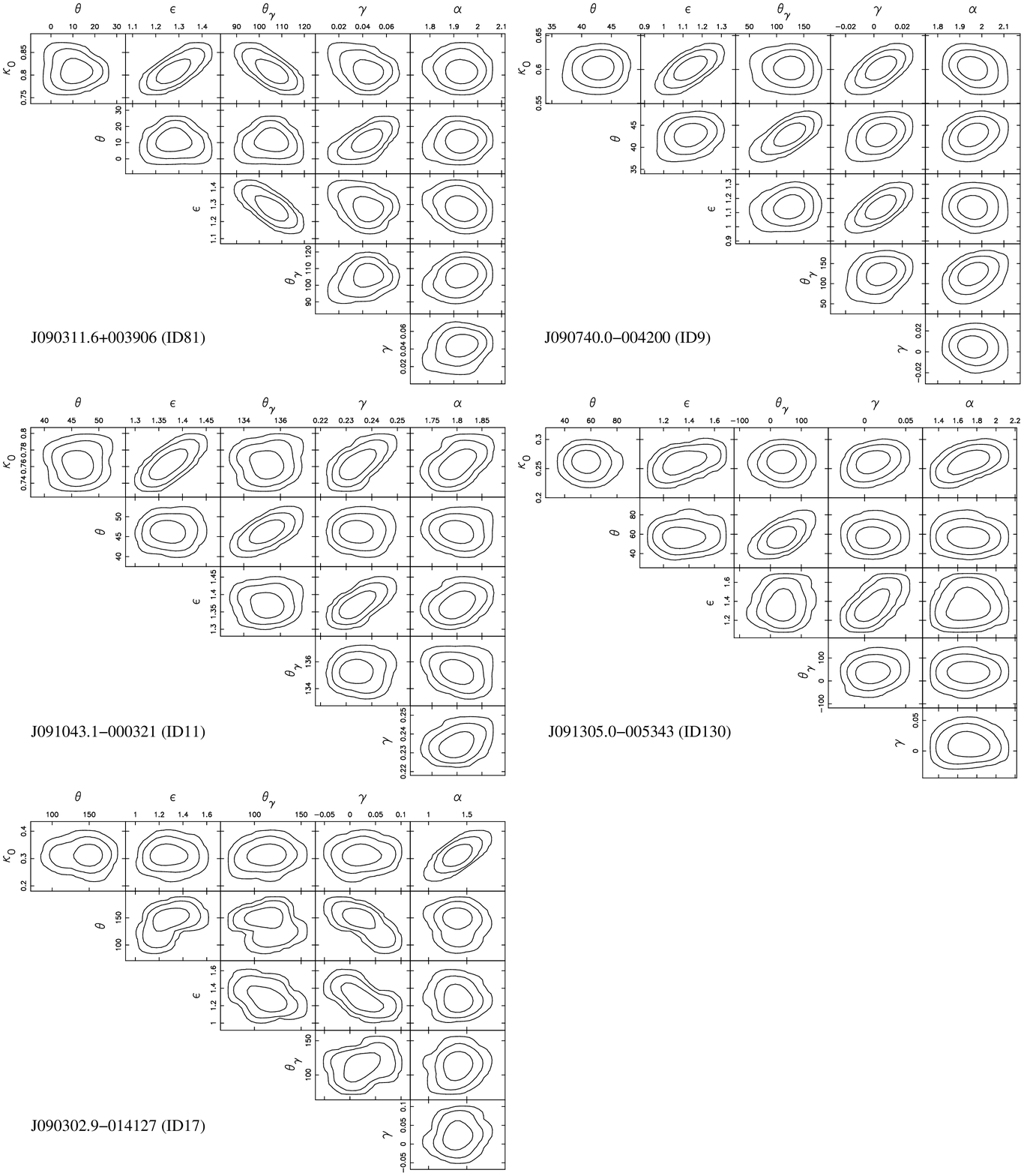}
\hfill}
\epsfverbosetrue
\caption{Parameter confidence limits. Contours show the 1, 2 and 3$\sigma$
single-parameter confidence regions for all parameter combinations,
excluding the position centroid of the lens.}
\label{params}
\end{figure*}

\section*{Acknowledgements}

The work in this paper is based on observations made with the NASA/ESA
Hubble Space Telescope under the HST programme \#12194. MN
acknowledges financial support from ASI/INAF Agreement I/072/09/0 and
from PRIN-INAF 2012 project: "Looking into the dust-obscured phase of
galaxy formation through cosmic zoom lenses in the Herschel
Astrophysical Large Area Survey". JGN acknowledges financial support
from the Spanish Ministerio de Ciencia e Innovacion, project
AYA2010-21766-C03-01, and the Spanish CSIC for a JAE-DOC fellowship,
co-funded by the European Social Fund.  We thank Martin Baes and
Michal Michalowski for constructive comments on this paper. We would
also like to thank Adam Moss for technical discussions regarding the
MCMC analysis contained herein.

\label{lastpage}

\end{document}